\documentclass[pre,epsfig,address,twocolumn]{revtex4}%

\usepackage{amsmath}
\usepackage{amssymb}
\usepackage{dsfont} 

\usepackage{graphicx}
\usepackage{epsfig}

\newcommand{\R}{{\mathbb R}}
\newcommand{\I}{{\mathbb I}}
\newcommand{\rmd}{\mathrm{d}}
\newcommand{\rme}{\mathrm{e}}
\newcommand{\sign}{\mathrm{sgn}}
\newcommand{\half}{\frac{1}{2}}
\newcommand{\bra}[1]{\mbox{$\langle \, {#1}\, |$}}
\newcommand{\ket}[1]{\mbox{$| \, {#1}\, \rangle$}}
\newcommand{\eref}[1]{(\ref{#1})}
\newcommand{\exval}[1]{\mbox{$\langle \, {#1}\, \rangle$}}
\newcommand{\bea}{\begin{eqnarray}}
\newcommand{\eea}{\end{eqnarray}}
\newcommand{\ba}{\begin{array}}
\newcommand{\ea}{\end{array}}
\newcommand{\bel}[1]{\begin{equation}\label{#1}}
\newcommand*{\cofrac}[2]{%
  {%
    \rlap{$\dfrac{1}{\phantom{#1}}$}%
    \genfrac{}{}{0pt}{0}{}{#1+#2}%
  }%
}

\newcommand{\be}{\begin{equation}}
\newcommand{\ee}{\end{equation}}

\begin{document}

\title{Fibonacci family of dynamical universality classes}

\author{Vladislav Popkov$^{1,2,3}$,   Andreas Schadschneider$^1$, Johannes Schmidt$^1$,
 and Gunter M.~Sch\"utz$^{4}$}
\affiliation{$^{1}$Institut f\"{u}r Theoretische Physik, Universit\"{a}t zu K\"{o}ln,
50937 Cologne, Germany.\\
$^{2}$ Centro Interdipartimentale per lo studio di Dinamiche Complesse,
 Universit\`a di Firenze,  50019 Sesto Fiorentino, Italy\\
$^3$ Present address: Helmholtz Institut f\"ur Strahlen und Kernphysik, University of Bonn,
Germany\\
$^4$ Theoretical
Soft Matter and Biophysics, Institute of Complex Systems II, Forschungszentrum J\"ulich, 52425 J\"ulich,
Germany\\
}
\begin{abstract}
Universality is a well-established central concept of equilibrium
  physics.  However, in systems far away from equilibrium a deeper
  understanding of its underlying principles is still lacking.  Up to
  now, a few classes have been identified.  Besides the diffusive
  universality class with dynamical exponent $z=2$ another prominent
  example is the superdiffusive Kardar-Parisi-Zhang (KPZ) class with
  $z=3/2$. It appears e.g. in low-dimensional dynamical phenomena
  far from thermal equilibrium which exhibit some conservation law.
  Here we show that both classes are only part of an infinite discrete
  family of non-equilibrium universality classes.  Remarkably their
  dynamical exponents $z_\alpha$ are given by ratios
  of neighbouring Fibonacci numbers, starting with either $z_1=3/2$ (if a
  KPZ mode exist) or $z_1=2$ (if a diffusive mode is present).  If
  neither a diffusive nor a KPZ mode are present, all dynamical modes
  have the Golden Mean $z=(1+\sqrt{5})/2$ as dynamical exponent. The
  universal scaling functions of these Fibonacci modes are asymmetric
  L\'evy distributions which are completely fixed by the macroscopic
  current-density relation and compressibility matrix of the system
  and hence accessible to experimental measurement.
\end{abstract}
\keywords{nonequilibrium physics | universality | dynamical exponent |
driven diffusion}

\maketitle



The Golden Mean $\varphi=1/2+\sqrt{5}/2\approx 1.61803..$,
also called Divine Proportion, has been an inspiring number for many
centuries. It is widespread in nature, i.e.\ arrangements of petals of
the flowers and seeds in the sunflower follow the golden rule
\cite{GoldenRatioLivio}.  Being considered an ideal proportion, the
Golden Mean appears in famous architectural ensembles such as
Parthenon in Greece, Giza Great Pyramids in Egypt, or Notre Dame de
Paris in France. Ideal proportions of the human body follow the Golden
Rule.

Mathematically, the beauty of the Golden Mean number is
expressed in its continued fraction representation: All the
coefficients in the representation are equal to unity,
\begin{equation}
\varphi=  1 +
  \cofrac{1}{
    \cofrac{1}{
      \cofrac{1}{
        \genfrac{}{}{0pt}{0}{}{\ddots}
  }}}
\label{DivineContiniousFraction}
\end{equation}
Systematic truncation of the above continued fraction gives the
so-called Kepler ratios, $1/1, 2/1,3/2,5/3,8/5,...$, which approximate
the Golden Mean. Subsets of denominators (or numerators) of the Kepler
ratios form the celebrated Fibonacci numbers, $F_i=1,1,2,3,5,8,..$, such
that Kepler ratios are ratios of two neighbouring Fibonacci numbers.
As well as the Golden Mean, Fibonacci ratios and Fibonacci numbers
are widespread in nature \cite{GoldenRatioLivio}.

The occurrence of the Golden Mean is not only interesting for
aesthetic reasons, but often indicates the existence of some
fundamental underlying structure or symmetry.  Here we demonstrate
that the Divine Proportion, as well as all the truncations (Kepler
ratios) of the continued fraction (\ref{DivineContiniousFraction}),
appear as universal numbers, viz., the dynamical exponents, in
low-dimensional dynamical phenomena far from thermal equilibrium. The
two well-known paradigmatic universality classes, Gaussian diffusion
with dynamical exponent $z=2$ \cite{Land14,Gend14} and the
Kardar-Parisi-Zhang (KPZ) universality class with $z=3/2$
\cite{Kard86} enter the Kepler ratios hierarchy as the
first two members of the family.

The universal dynamical exponents in the present context characterize
the self-similar space-time fluctuations of locally conserved
quantities, characterizing e.g. mass, momentum or thermal transport in
one-dimensional systems far from thermal equilibrium \cite{Spoh14}.
The theory of nonlinear fluctuating hydrodynamics (NLFH), has recently
emerged as a powerful and versatile tool to study space-time
fluctuations, and specifically the dynamical structure function which
describes the behaviour of the slow relaxation modes, and from which
the dynamical exponents can be extracted \cite{Spohn15}.

The KPZ universality class \cite{KPZniceIntroduction}
 has
been shown to explain the dynamical exponent observed in interface
growth processes as diverse as the propagation of flame fronts
\cite{Maun97,Miet05}, the growth of bacterial colonies \cite{Waki97},
or the time evolution of droplet shapes such as coffee stains
\cite{Yunk13} where the Gaussian theory fails.  The dynamical
structure function originating from the one-dimensional KPZ equation
has a non-trivial scaling function obtained exactly by Pr\"ahofer and
Spohn from the totally asymmetric simple exclusion process (TASEP) and
the polynuclear growth model \cite{Prae02,Prae04} and was beautifully
observed in experiments on turbulent liquid crystals
\cite{Take10,Take11}. Theoretical treatment, both numerical
and analytical, of generic model systems with Hamiltonian dynamics
\cite{vanB12}, anharmonic chains \cite{Mend13,Spoh14b} and lattice
models for driven diffusive systems \cite{Popk14,Popk14b}, have
demonstrated an extraordinary robust universality of fluctuations of
the conserved slow modes in one-dimensional systems.

Despite this apparent ubiquity,  dynamical exponents
different from $z=2$ or $z=3/2$ were observed frequently. Usually it is not clear
whether this corresponds to genuinely different dynamical critical
behaviour or is just a consequence of imperfections in the
experimental setting.  Moreover, recently a new universality class
with dynamical exponent $z=5/3$ for the heat mode in Hamiltonian
dynamics \cite{vanB12} was discovered, followed by the discovery of
some more universality classes in anharmonic chains
\cite{Mend13,Spoh14b} and lattice models for driven diffusive systems
\cite{Popk14,Popk14b}.  What is lacking, even in the conceptually
simplest case of the effectively one-dimensional systems that we are
considering, is the understanding of the plethora of dynamical
non-equilibrium universality classes within a larger framework. Such a
framework exists e.g. for two-dimensional critical phenomena in
equilibrium systems where the spatial symmetry of conformal invariance
together with internal symmetries give rise to discrete families of
universality classes in which all critical exponents are simple
rational numbers.

It is the aim of this article to demonstrate that  discrete families of universality classes with
fractional critical exponents appear also far from
thermal equilibrium. This turns out to be a hidden
feature of the NLFH equations which we extract using mode-coupling
theory. It is remarkable that one finds dynamical exponents $z_\alpha$
which are ratios of neighbouring Fibonacci numbers
$\{1,1,2,3,5,8,\dots\}$ defined recursively as
$F_{n}=F_{n-1}+F_{n-2}$. The first two members of this family are
diffusion ($z=2=F_3/F_2$) and KPZ ($z=3/2=F_4/F_3$).  The
corresponding universal scaling functions are computed and shown to be
(in general asymmetric) $z_\alpha$-stable L\'evy distributions with
parameters that can be computed from the macroscopic current-density
relation and compressibility matrix of the corresponding physical
system and which thus can be obtained from experiments without
detailed knowledge of the microscopic properties of the system.  The
theoretical predictions, obtained by mode coupling theory, are
confirmed by Monte-Carlo simulations of a three-lane asymmetric simple
exclusion process which is a model of driven diffusive transport of
three conserved particle species.


\section{Nonlinear fluctuating hydrodynamics}

We consider a rather general interacting non-equilibrium system of
length $L$ described macroscopically by $n$ conserved order parameters
$\rho_\lambda(x,t)$ with stationary values $\rho_\lambda$ and
associated macroscopic stationary currents
$j_\lambda(\rho_1,\dots,\rho_n)$ and compressibility matrix $K$ with
matrix elements $K_{\lambda \mu} = \frac{1}{L} \left\langle(N_\lambda
  - \rho_\lambda L) (N_\mu - \rho_\mu L)\right\rangle$ where
$N_\lambda = \int_0^L \rmd x \rho_\lambda(x,t)$ are the
time-independent conserved quantities.

The starting point for investigating density fluctuations $u_\lambda(x,t) :=
\rho_\lambda(x,t) - \rho_\lambda$ in the non-equilibrium steady state
are the NLFH equations \cite{Spoh14}
\begin{equation}
\label{coupledBurgers}
\partial_t \vec{u} =  - \partial_x \left( J \vec{u} +
\frac{1}{2} \langle u|  \vec{H} |u\rangle - \partial_x  D \vec{u}
+ B \vec{\xi} \right)
\end{equation}
where $J$ is the current Jacobian with matrix elements $ J_{\lambda
  \mu} = \partial j_\lambda / \partial \rho_\mu$, $\vec{H}$ is a
column vector whose entries $\left(\vec{H}\right)_\lambda=H^{\lambda}$
are the Hessians with matrix elements $H^{\lambda}_{\mu \nu} =
\partial^2 j_\lambda /(\partial \rho_\mu \partial \rho_\nu)$ and the
bra-ket notation represents the inner product in component space
$\bra{u} (\vec{H})_\lambda \ket{u} = \vec{u}^T H^{\lambda} \vec{u} =
\sum_{\mu \nu} u_\mu u_\gamma H^{\lambda}_{\mu \nu}$ with $\bra{u} =
\vec{u}^T$, $\ket{u} = \vec{u}$.  The diffusion matrix $D$ is a
phenomenological quantity.  The noise term $B\vec{\xi}$ does not appear
explicitly below, but plays an indirect role in the mode-coupling
analysis.  The product $JK$ of the Jacobian with the compressibility
matrix $K$ is symmetric \cite{Gris11} which guarantees a
hyperbolic system of conservation laws \cite{Toth03}.  We ignore
possible logarithmic corrections arising from cubic contributions
\cite{Devi92}.

This system of coupled noisy Burgers equations is conveniently treated
in terms of normal modes $\vec{\phi}=R \vec{u}$ where $ RJR^{-1} =
\mathrm{diag}(v_\alpha)$ and the transformation matrix $R$ is
normalized such that $RKR^T = 1$. The eigenvalues $v_\alpha$ of $J$
are the characteristic velocities of the system.  From
\eref{coupledBurgers} one thus arrives at
$\partial_t \phi_\alpha = -  \partial_x \left( v_\alpha \phi_\alpha +
\langle  \phi| G^{\alpha} |\phi \rangle - \partial_x  (\tilde{D}
\vec{\phi})_\alpha + (\tilde{B} \vec{\xi})_\alpha \right)$
with $\tilde{D}=RDR^{-1}$, $\tilde{B}=RB$ and the mode coupling matrices
\begin{equation}
G^{\alpha} =  \frac{1}{2} \sum_\beta R_{\alpha \beta} (R^{-1})^T
H^{\beta} R^{-1}
\end{equation}
whose matrix elements we denote by $G^\alpha_{\beta \gamma}$.\\


\section{Computation of the dynamical structure function}

The dynamical structure function describes the
stationary fluctuations of the conserved slow modes and is thus a key
ingredient for understanding the interplay of noise and non-linearity
and their role for transport far from equilibrium.  We focus on the
case of strict hyperbolicity where all $v_\alpha$ are pairwise
different and study the large scale behaviour of the dynamical
structure function $S^{\alpha \beta}(x,t) =
\exval{\phi_\alpha(x,t)\phi_\beta(0,0)}$.  Since all modes have
different velocities only the diagonal elements $S_\alpha(x,t) :=
S^{\alpha\alpha}(x,t)$ are non-zero for large times.  Mode coupling
theory yields \cite{Spoh14}
\bea
\label{modecoupling}
& &\partial_t S_\alpha(x,t) =  (-v_\alpha \partial_x +
D_\alpha \partial_x^2) S_\alpha(x,t) \nonumber \\
 &+ &\int_0^t \rmd s \int_{\R}
\rmd y S_\alpha(x-y,t-s) \partial_y^2 M_{\alpha\alpha}(y,s)
\eea
with the diagonal element $D_\alpha:=\tilde{D}_{\alpha\alpha}$ of the
phenomenological diffusion matrix for the eigenmodes and the memory kernel
$M_{\alpha\alpha}(y,s) = 2 \sum_{\beta,\gamma} (G^{\alpha}_{\beta \gamma})^2
S_\beta(y,s)S_\gamma(y,s).$
The task therefore is to extract for arbitrary $n$ the large-time and
large-distance behaviour from this non-linear integro-differential
equation.

Remarkably, these equations can be solved exactly in the
long-wavelength limit and for $t \rightarrow \infty$ by Fourier and
Laplace transformation (see Appedices).  Using a suitable scaling ansatz for the transformed
structure function then allows to analyze the small-$p$ behaviour from
which the dynamical exponents can be determined.  We find that
different conditions arise depending on which diagonal elements of the
mode-coupling matrices vanish.

\section{The Fibonacci family of dynamical universality classes}

\subsection{Fibonacci case}

First, we consider the case where the self-coupling
$G^{\alpha}_{\alpha\alpha}$ is nonzero for one mode only, e.g.
$G^{1}_{11} \neq 0$. For all other modes $\alpha > 1$ we assume a
single nonzero coupling to the previous mode, so
$G^{\alpha}_{\alpha-1, \alpha-1} \neq 0$, and $G^{\alpha}_{\beta,
  \beta}=0$ for $\beta\neq \alpha-1$.  Then, as follows from our
analysis (see Appendices),
we find the following recursion for the dynamical exponents:
\bel{RecurrenceExponents}
z_\alpha = 1 + \frac{1}{z_{\alpha-1}}
\ee
with $z_1 = 3/2$.

The dynamical structure function in momentum space is proportional to
the $z_\alpha$-stable L\'evy distribution with maximal asymmetry
$\sigma^\alpha=\pm 1$, see \cite{Durrett} and
  eq.~(\ref{Scalingfunction}) below.  The sign of the asymmetry
depends whether the mode $(\alpha-1)$ has bigger or smaller velocity
than the mode $\alpha$, $\sigma^\alpha=-\sign(v_\alpha-v_{\alpha-1})$.
The dynamical exponents \eref{RecurrenceExponents} form a sequence of
rational numbers \bel{Fiboseq} z_\alpha =
\frac{F_{\alpha+3}}{F_{\alpha+2}} \ee which are consecutive ratios of
neighbouring Fibonacci numbers $F_\alpha$, defined by $F_\alpha =
F_{\alpha-1} + F_{\alpha-2}$ with initial values $F_0=0,\quad F_1=1$,
which converge exponentially to the Golden Mean $\varphi := \half
(1+\sqrt{5}) \approx 1.618$, as first observed by Kepler in 1611 in a
treatise on snow flakes.  In a model with $n$ conservation laws, one
has the Fibonacci modes with dynamical exponents $\{3/2,\, 5/3,\,
8/5,...,\, z_n\}$.

Finally, we remark that if mode 1 is diffusive rather than KPZ, then
we find the same sequence (\ref{Fiboseq}) of exponents except that it
starts with $z_1=F_2=2$.

\begin{figure}[ptb]
\centerline{
\includegraphics[width=0.5\textwidth]{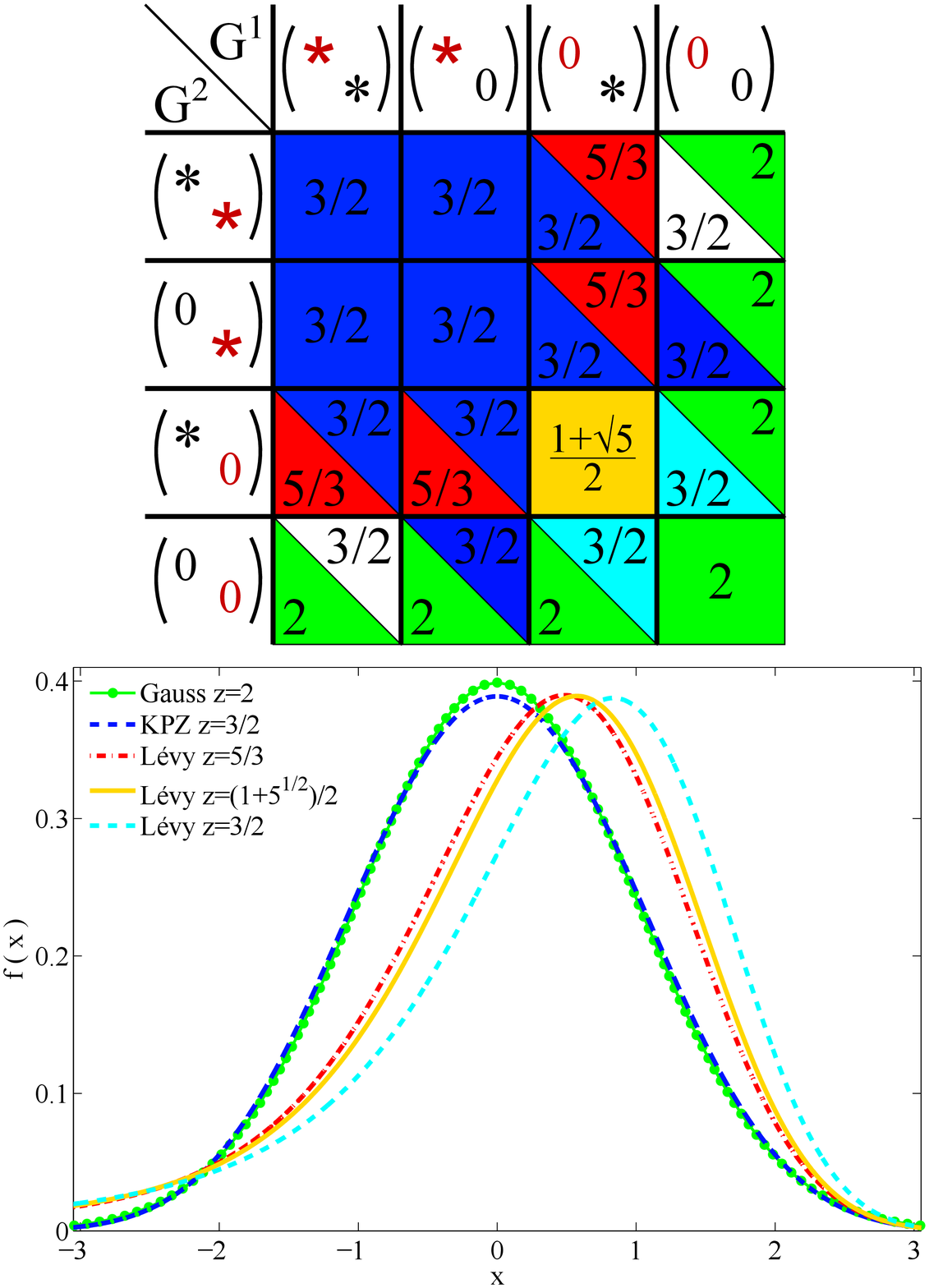}
}
\caption{The scaling functions (bottom) and dynamical exponents
    are related to the structure of the mode coupling matrices
    $G^\alpha$ (top). The table shows the dynamical exponents
    $z_\alpha$ in the case $n=2$, see eq.~(\ref{dynexpsumm2}). The
    symbols $\ast$ and $\star$ denote non-zero elements. Red symbols
    correspond to self-coupling, black symbols to couplings to other
    modes.  Matrix elements not indicated can take any value. The
    colors in the table correspond to the colors of the graphs of the
    scaling functions.
\label{FigScalingFct}
}
\end{figure}

In Fig.~\ref{FigScalingFct} we show some representative examples
of the scaling functions which are quite different in shape.
Furthermore the relation between the exponents $z_\alpha$ , determined by
eq.~(\ref{dynexpsumm2}), and the mode
coupling matrices $G^\alpha$ is illustrated for the case $n=2$.

\subsection{Golden Mean case}

As second representative example we consider the case where all
self-coupling coefficients vanish, $G^{\alpha}_{\alpha\alpha}\equiv 0$
for all $\alpha$, while each mode has at least one nonzero coupling to
another mode, $G^{\alpha}_{\beta\beta}\neq 0$ for some $\beta\neq
\alpha$.  Then, \eref{RecurrenceExponents} reduces to $z_\alpha = 1 +
1/z_\beta$ for all modes $\alpha,\ \beta$. The unique solution of this
equation is the Golden Mean $z_\alpha = \varphi= (1+\sqrt{5})/2$ for
all $\alpha$. The scaling functions (see Supporting Information section)
are proportional to $\varphi$-stable L\'evy distributions with
parameters fixed by the collective velocities and the mode-coupling
coefficients.  The asymmetry of the fastest right-moving (left-moving)
mode is predicted to be $\beta=-1$ ($\beta=1$).


\section{Simulation results}

\begin{figure}[ptb]
\centerline{
\includegraphics[width=0.45\textwidth]{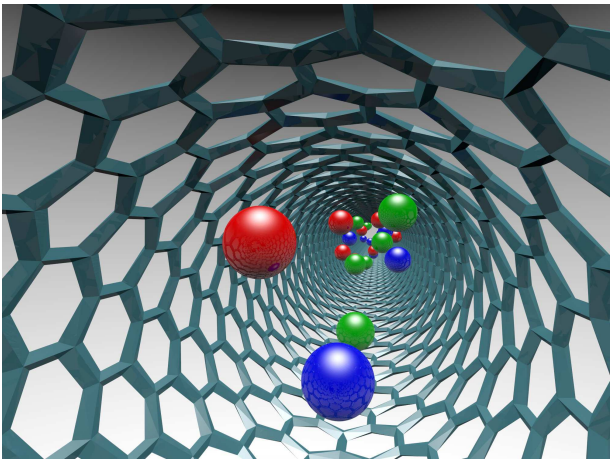}
}
\caption{Schematic drawing of three particle species drifting inside a
  nanotube.  Due to the interaction between the particles and with the
  walls one expects a non-linear current density relation.
\label{FigTube}
}
\end{figure}

To check the theoretical predictions for the two cases we
simulate mass transport with three conservation laws, i.e., three
distinct species of particles. To maintain a far-from-equilibrium
situation a driving force is applied that leads to a constant drift
superimposed on undirected diffusive motion. This is a natural setting
for transport of charged particles in nanotubes, see Fig.~\ref{FigTube} for an
illustration, where a direct
measurement of the stationary particle currents is experimentally
possible \cite{Lee10}.  However, due to the universal applicability of
NLFH the actual details of the interaction of the particles with their
environment and the driving field are irrelevant for the theoretical
description of the large-scale dynamics. Hence for good statistics we
simulate a lattice model for transport which represents a minimal
realization of the essential ingredients, namely a non-linear
current-density relation for all three conserved masses.

Our model is the three-species version of the multi-lane totally
asymmetric simple exclusion process \cite{Popk04}. Particles hop
randomly in field direction on three lanes to their neighbouring sites
on a periodic lattice of $3\times L$ sites with rates that depend on
the nearest-neighbour sites. Lane changes are not allowed so that
the total number of particles on each lane is conserved.
Due to excluded-volume interaction each
lattice site can be occupied by at most one particle. Thus the
occupation numbers $n_k^{(\lambda)}$ of site $k$ on lane $\lambda$
take only values 0 or 1. The hopping rate $r_k^{(\lambda)}$ from site
$k$ on lane $\lambda$ to site $k+1$ on the same lane is given by
\bel{hoppingrates}
r_k^{(\lambda)} = b_\lambda + \half \sum_{\{\mu :\, \mu\neq \lambda\}}
\gamma_{\lambda\mu} \left(n_k^{(\mu)} + n_{k+1}^{(\mu)}\right)
\ee
with a species dependent drift parameter $b_\lambda$ and symmetric
interaction constants $\gamma_{\lambda\mu} = \gamma_{\mu\lambda}$.
Hopping attempts onto occupied sites are rejected.  The conserved
quantities are the three total numbers of particles $N_\lambda$ on
each lane with corresponding densities $\rho_\lambda=N_\lambda/L$.

The stationary distribution of our model factorizes \cite{Popk04} and
thus allows for the exact computation of the macroscopic
current-density relations $j_\lambda(\rho_1,\rho_2,\rho_3)$ and the
compressibility matrix $K(\rho_1,\rho_2,\rho_3)$. Furthermore, because  there is no
particle exchange between lanes, the compressibility matrix is diagonal
with elements denoted by $\kappa_\lambda$.  One has
\bea
\label{curr}
j_\lambda & = & \rho_\lambda(1-\rho_\lambda)\left(b_\lambda +
\sum_{\{\mu :\, \mu\neq \lambda\}} \gamma_{\lambda \mu} \rho_\mu\right) \\
\label{comp}
\kappa_\lambda & = & \rho_\lambda(1-\rho_\lambda).
\eea
The diagonalization matrix $R$ and the mode-coupling matrices $G^\alpha$ are
fully determined by these quantities.

\begin{figure}[ptb]
\centerline{
\includegraphics[width=0.5\textwidth]{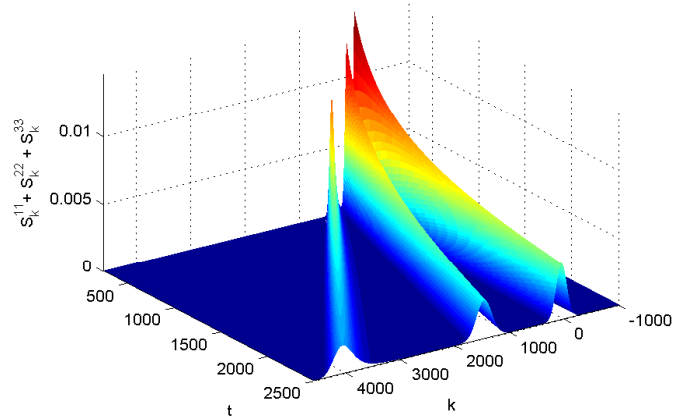}}
\caption{Space-time propagation of three normal modes in the three-lane
  model.  The modes (from left to right) are the Fibonacci mode with
  $z=8/5$ (mode 3), the KPZ mode with $z=3/2$ (mode 1), and the
  Fibonacci mode with $z=5/3$ (mode 2).  The physical and
    simulation parameters are given in the Appendices.}
\label{Fig3Dpeaks}
\end{figure}

\begin{figure}[ptb]
\centerline{
\includegraphics[width=0.5\textwidth]{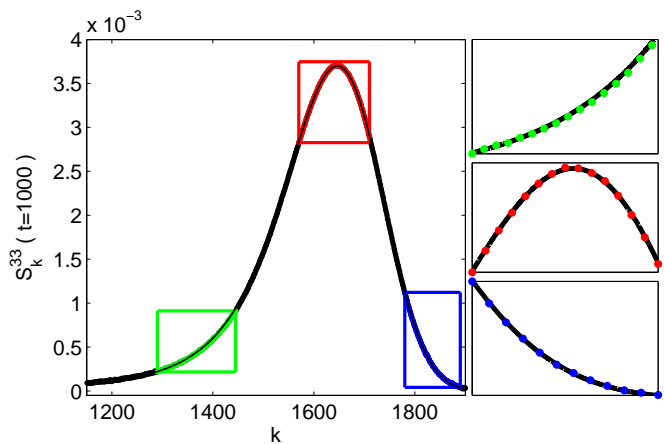}
}
\caption{Left Panel: Vertical least squares fit of the numerically obtained
dynamical structure function for the Fibonacci 8/5-mode (points),
at time $t=1000$ with a 8/5-stable L\'evy distribution,
maximal asymmetry $-1$ and theoretical center of mass (line), predicted
by the mode coupling theory. The only fit parameter is the scale parameter
of the L\'evy stable distribution.
The simulation results agree very well with the asymptotic
theoretical result already for moderate times.
Right Panel: Insets show closeups of the peak region and tail regions,
according to a colour code. Every $10$-th datapoint is plotted to
improve the visibility of the data. The statistical error
$\epsilon_{99\%}$ with 99\% confidence bound is for every data point
smaller than $1.6299\cdot10^{-5}$.
\label{FigLeviFit}
}
\end{figure}

According to mode-coupling theory three different Fibonacci-modes with
$z_1 = 3/2, z_2 = 5/3, z_3 = 8/5$ occur e.g. when $G^1_{11} \neq 0$,
$G^2_{11} \neq 0$, $G^3_{22} \neq 0$, and $G^2_{22} = G^2_{33} =
G^3_{33} = 0$. For our simulation we compute numerically densities,
bare hopping rates and interaction parameters to satisfy these
properties as described in the Appendices.  For this
choice the velocities of the normal modes are $v_{1}=0.592315$,
$v_2= 0.0281578$, $v_3=1.58226$ which ensures a good spatial
separation after quite small times.  The propagation of the three
normal modes (Fig.~\ref{Fig3Dpeaks}) with the predicted velocities is
observed with an error of less than $10^{-3}$. Moreover, the
numerically obtained dynamical structure function for mode $3$ shows a
startling agreement with the theoretically predicted L\'evy scaling
function with $z=8/5$ and maximal asymmetry, see
Fig.~\ref{FigLeviFit}. It takes longer for the other two modes (KPZ
mode and L\'evy stable $5/3$ mode) to reach their asymptotic form,
which we argue is due to the much smaller respective couplings,
$(G^1_{11}/G^3_{22})^2 \ll 1$, $( G^2_{11}/G^3_{22})^2 \ll 1$.

\begin{figure}[ptb]
\centerline{
\includegraphics[width=0.5\textwidth]{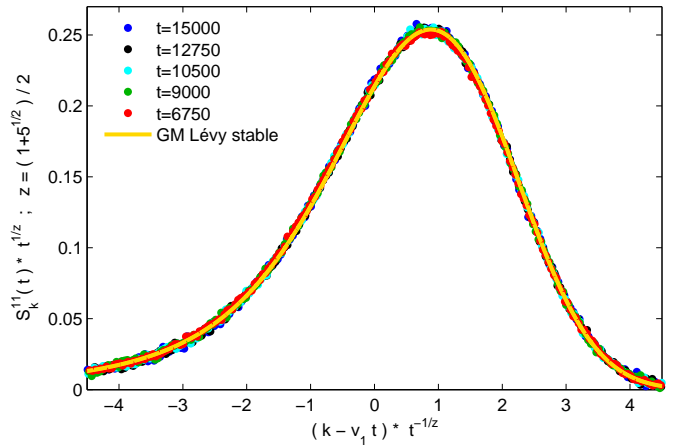}}
\caption{Scaling plot of the measured structure function
    of mode $1$ with dynamical exponent
    $z\equiv\varphi=\left(1+\sqrt{5}\right)/2$ for the Golden Mean
    case, fitted to a $\varphi$-stable L\'evy distribution with maximal
    asymmetry $-1$ (see Eq.~(\ref{Scalingfunction})).  The scale parameter
    $E_{1}$ for the L\'evy-stable distribution and the center of mass
    velocity $v_{1}$ are obtained by a vertical least square fit.
    Fitted parameters are $v_{1,\text{fit}}=1.83107\pm0.00009$ and
    $E_{1,\text{fit}}=1.1\pm0.01$.  The fitted velocity
    $v_{1,\text{fit}}$ differs by $0.02\%$ from the theoretical
    velocity.
\label{FigScaling}
}
\end{figure}

In order to observe three Golden Mean modes it is sufficient to
require that each mode has zero self-coupling and at least one nonzero
coupling to other modes. This can be achieved with the set of
parameters given in Appendices which lead to the
velocities $v_1=1.83149$, $v_2= 0.762688$, $v_3 =0.326778$ of the
normal modes.  The propagation of the three normal modes with the
predicted velocities is observed, approaching for large times a very
small relative error of about $10^{-4}$.  The structure function for
the fastest mode $1$ converges to its asymptotic shape faster than for
the other modes, due to the large coupling coefficient $G^1_{33}$.  In
Fig.~\ref{FigScaling} we show a scaling plot of the measured structure
function for mode $1$ with dynamical exponent
$z\equiv\varphi=\left(1+\sqrt{5}\right)/2$ together with a fitted to a
$\varphi$-stable L\'evy function \eref{Scalingfunction} with maximal
asymmetry $\beta=-1$ as predicted by the theory.  The data collapse
shows a striking agreement between the measured and theoretical
scaling function.  Alternatively, the dynamical exponent $z_\alpha$
can be derived from the maximum of the structure function, which
scales as $\max(S_1(x,t))=\text{const}\cdot t^{-1/z}$.  We
obtain $z\approx 1.63$ which differs from the predicted value $z
\equiv \varphi=(1+\sqrt 5)/2$ by less than $0.8\%$.\\

\section{Discussion}

Our work demonstrates that non-equilibrium phenomena are much richer
than just the diffusive and KPZ universality suggest.  We have established
that in non-equilibrium phenomena governed by non-linear fluctuating
hydrodynamics with $n$ conservation laws mode coupling theory predicts
a family of dynamical universality classes with dynamical exponents
given by the sequence of consecutive Kepler ratios \eref{Fiboseq} of
Fibonacci numbers.  With slightly modified initial conditions
on $G_1^{11}$ this result is easily generalized for the case
when the first mode $\alpha=1$ is diffusive. Then the sequence of
dynamical exponents becomes shifted by one unit with respect to
(\ref{Fiboseq}). On the other hand, if all self-couplings vanish,
but at least one other diagonal element $G^\alpha_{\beta\beta}$
of the mode coupling matrix is non-zero,one
has as unique solution for all modes $\alpha$ the fixed point value
$z_\alpha = z_\infty = \varphi$ which is the Golden Mean.

For general mode coupling matrices all critical exponents can be
computed (from \eref{dynexpsumm2} in Appendices).  The
scaling functions of the non-diffusive and non-KPZ modes are
asymmetric L\'evy distributions whose parameters are completely
determined by the macroscopic current-density relation and
compressibility matrix of the system.

For 1+1 dimensional systems out of equilibrium this is the first time, to our knowledge, that
an infinite family of discrete universality classes is found.
Recalling that 1+1 dimensional non-equilibrium systems with
short-range interactions can be mapped onto two-dimensional equilibrium
systems (with the time evolution operator playing the role of the
transfer matrix) one is reminded of the discrete families of
conformally invariant critical equilibrium systems in two space
dimensions \cite{Card96,Henk99}. We do not know whether there is any
mathematical link, but the analogy is suggestive in so far as
conformal invariance is a local symmetry of spatially isotropic
systems with $z=1$ (which happens to be the lowest order Kepler ratio)
while $z>1$ corresponds to strongly anisotropic systems for which also
local symmetry groups are known to exist \cite{Henk02}.

Since an infinite number of lanes of coupled one-dimensional systems
corresponds to a two-dimensional system, it is intriguing to observe
that the Golden Mean is close to the numerical value $z=1.612-1.618$
of the dynamical exponent of the $2+1$-dimensional KPZ-equation
\cite{Parisi,HalpinHealy2}. The scaling function of the
$2+1$-dimensional KPZ-equation, however, is not L\'evy
\cite{Scaling2dKPZ}.

In order to observe and distinguish between the different new classes
highly precise experimental data will be required.  E.g.\ in the
Fibonacci case the dynamical exponents converge quickly to the Golden
Mean.  A feature which might be easier to observe experimentally is
the scaling function itself, which for higher Fibonacci ratios
$5/3,8/5,...$ usually has a strong asymmetry (see
Figs.~\ref{FigScalingFct}, \ref{FigLeviFit} and \ref{FigScaling})
while KPZ and Gauss scaling functions are symmetric.  Growth processes
which can be mapped on exclusion processes with several conservation
laws, might be potentially suitable candidates for an experimental
verification, see e.g.\ \cite{Take10,Take11} for an example of a
system with one conservation law.



\begin{appendix}

\section{Computation of the dynamical structure function}
The mode coupling equations (\ref{modecoupling}) can be solved
in the scaling limit by applying a Fourier transform (FT) $f(x) \to
\tilde f(p)$ and a Laplace transform (LT) $f(t) \to \tilde f(\omega)$.
For more details we refer to \cite{Popk14b} where the case $n=2$
of two conservation laws has been treated.
After making the scaling ansatz
\bel{ScaleFLT}
\tilde{S}_\alpha(p,\tilde{\omega}_\alpha) = p^{-z_\alpha}
g_\alpha(\zeta_\alpha)
\ee
for the transformed dynamical structure function where
$\hat{S}_\alpha(p,0)= 1/\sqrt{2\pi}$ and $\zeta_\alpha =
\tilde{\omega}_\alpha |p|^{-z_\alpha}$ we are in a position to analyze
the small-$p$ behaviour.  One has to search for dynamical exponents
for which the limit $p\to 0$ is non-trivial, which requires a
self-consistent treatment of all modes. We find that different
conditions arise depending on which diagonal elements of the
mode-coupling matrices vanish.  To characterize the possible
scenarios we define the set $\I_\alpha := \{\beta:
G^{\alpha}_{\beta\beta} \neq 0\}$ of non-zero diagonal mode coupling
coefficients. Through power counting one obtains
\be
\label{dynexpsumm2}
z_\alpha = \left\{ \ba{lll}
2 & \mbox{ if } & \I_\alpha = \emptyset  \\
3/2 & \mbox{ if } & \alpha \in \I_\alpha\\
\min_{\beta \in \I_\alpha}\left[\left(1 + \frac{1}{z_\beta}\right)\right]
& \mbox{ else } &
\ea \right.
\ee
and the domain
\be
\label{dynexpsumm1}
1 < z_\alpha \leq 2 \quad \forall \alpha.
\ee
for the possible dynamical exponents.\\

In the Fibonacci case, the dynamical structure function of mode
$\alpha$ in momentum space has the scaling form
\bel{Scalingfunction}
\hat{S}_\alpha(p,t) = \frac{1}{\sqrt{2\pi}} \rme^{-iv_\alpha
  pt-E_\alpha |p|^{z_\alpha}t \left(1 - i
    \sigma_p^{\alpha\beta}\tan{\left(\frac{\pi z_\alpha}{2}\right)}
  \right)}
\ee
with inverse time scales $E_\alpha$. The dynamical exponents then
satisfy the recursion (\ref{RecurrenceExponents}).
Up to the normalization $1/\sqrt{2\pi}$ the scaling form
(\ref{Scalingfunction}) is a $\alpha$-stable L\'evy distribution
\cite{Durrett}.\\


\section{Simulation algorithm}
For the Monte-Carlo simulation of the model we choose a large system
size $L \geq 5\cdot 10^5$ which avoids finite-size effects. At time $t=0$,
$N_\lambda$ particles are placed on each lane according to the desired
initial state. One Monte-Carlo time unit consists of $3\cdot L\cdot r^*$
random sequential update steps where $r^* = \max\left\{{r_{k}}^{(\lambda)}
\right\}$: In each update step a bond
$(k^{(\lambda)},k^{(\lambda)}+1)$ is chosen randomly with uniform
distribution.  If $n_k^{(\lambda)}\left(1-n_{k+1}^{(\lambda)}\right)=1$
then the particle at site $k$ is moved to $k+1$ with probability
$r_k^{(\lambda)}/r^\ast$ where $r^\ast$ is the maximal value that the
$r_k^{(\lambda)}$ can take among all possible particle configurations
on the neighbouring lanes. If
$n_k^{(\lambda)}(1-n_{k+1}^{(\lambda)})=0$ the particle configuration
remains unchanged.


\section{Simulation of the dynamical structure function}
In order to determine the dynamical structure function we initialize
the system by placing $N_\lambda$ particles uniformly on each lane
$\lambda$. This yields a random initial distribution drawn from the
stationary distribution of the process.  No relaxation is required.

Then we use translation invariance and compute the space- and time
average \be \sigma_{L,k}^{\lambda \mu}(M,\tau,t) = \frac{1}{M}
\sum_{j=1}^M \frac{1}{L} \sum_{l=1}^L
n_{l+k}^{(\lambda)}(j\tau+t)n_l^{(\mu)}(j\tau) - \rho_\lambda\rho_\mu.
\label{eq-sigma}
\ee
To avoid noisy data of $\sigma_{L,k}^{\lambda \mu}$
we take in (\ref{eq-sigma}) the system size $L$ and the time average parameter $M$ sufficiently
large. In order to obtain $S_{k}^{\lambda \mu}(t)$ we average over
$P$ independently generated and propagated initial configurations
of $\sigma_{L,k}^{\lambda \mu}$. The error estimates for
$S_{k}^{\lambda \mu}(t)$ are calculated from the $P$ independent
measurements.
From $S^{\lambda \mu}_k(t)$ we compute the structure function of the
normal modes by transformation with the diagonalizing matrix $R$
determined by \eref{curr} and \eref{comp}.

To obtain model parameters for three different
Fibonacci-modes with $z_1 = 3/2, z_2 = 5/3, z_3 = 8/5$ we solve the
equations given in the text after Eq.~\ref{comp} numerically with a
C-program performing direct minimization of the absolute values of the
targeted G-elements until the given tolerance value ($10^{-6}$) is
reached.  The data shown here for the three mode case have been
obtained from simulations with densities $\rho_1=0.2,\, \rho_2=0.25,\,
\rho_1=0.3$, bare hopping rates $b_1=0.613185$, $b_2= 0.425714$, $b_3=
0.799831$ and interaction parameters $\gamma_{12}=1.36145$,
$\gamma_{23}=3.69786$, $\gamma_{13}=0.143082$ for which the needed
relations are satisfied.  This choice of parameters yields
$G^1_{11}=0.322507$, $G^2_{11}= -0.15$, $G^3_{22}= 1.04547$, while the
the absolute values of $G^2_{22}, \, G^2_{33},\, G^3_{33}$ are smaller
than $10^{-6}$.  Besides these physical parameters, the simulation
parameters for the Fibonacci modes (Fig.~\ref{Fig3Dpeaks},
\ref{FigLeviFit}) are $L=5\cdot 10^5$, $\tau=250$, $M=1400$, $P=98$.

For the Golden Mean case (Fig.~\ref{FigScaling}) the set of
parameters $\rho_1=0.2,\ \rho_2=\rho_3=0.25$,
$\gamma_{12}=0.0082334758646$, $\gamma_{23}= 1.68447706968$,
$\gamma_{13} = 3.72140740146$, and $b_1=0.905073261248$, $b_2= 0.86$,
and $b_3= 1.18875738638$. This leads to $G^1_{22}=0.405702$,
$G^1_{33}= 0.929315$, $G^2_{11}=-0.104141$, $G^2_{33}= -0.208477$,
$G^3_{11}=-0.182467$, $G^3_{22}= 0.271246$, while the absolute values
of $G^1_{11}, G^2_{22},G^3_{33}$ are smaller than $10^{-6}$.  The
simulation parameters for the Golden Mean case are $L=5\cdot 10^6$,
$\tau=750$, $M=30$ and $P=303$.



\section*{Supporting Information}

Remarkably, the mode coupling equations eq.~(\ref{modecoupling})   can be
solved exactly in the scaling limit by Fourier and Laplace transformation.
To this end we define the Fourier transform (FT) as
\bel{FT}
\hat{f}(p) := \frac{1}{\sqrt{2\pi}} \int_{-\infty}^\infty \rmd x\,
\rme^{-ipx} f(x),
\ee
and the Laplace transform (LT) as
\bel{LT}
\tilde{f}(\omega) :=  \int_{0}^\infty \rmd t\, \rme^{-\omega t} f(t).
\ee
With $\widehat{D}_\alpha(p) = i v_\alpha p + D_\alpha p^2$ we obtain
from eq.~(4) of the paper in momentum-frequency space
\bel{ModecouplingFLT}
\tilde{S}_\alpha(p,\omega) = \frac{\hat{S}_\alpha(p,0)}{\omega
+ \widehat{D}_\alpha(p) +  \tilde{C}_{\alpha\alpha}(p,\omega) }
\ee
with memory kernel
\be
\label{memoryFLT}
\tilde{C}_{\alpha\alpha}(p,\omega) = 2 \sum_{\beta,\gamma}
(G^{\alpha}_{\beta\gamma})^2 p^2
\int_{0}^\infty \rmd s\, \rme^{-\omega s}
\int_{\R}
\rmd q \hat{S}_\beta(q,s) \hat{S}_\gamma(p-q,s).
\ee
and $\hat{S}_\alpha(p,0)= 1/\sqrt{2\pi}$.

Next we introduce $\tilde{\omega}_\alpha := \omega + i v_\alpha p$
and make the scaling ansatz
\bel{ScaleFLTa}
\tilde{S}_\alpha(p,\tilde{\omega}_\alpha) = p^{-z_\alpha}
g_\alpha(\zeta_\alpha)
\ee
with $\zeta_\alpha = \tilde{\omega}_\alpha |p|^{-z_\alpha}$. Having in
mind systems with short-range interactions we anticipate that all
modes spread subballistically, i.e., $z_\alpha > 1$ for all $\alpha$.
Using strict hyperbolicity one obtains after some substitutions of
variables

\bea
\label{memoryLFT5}
g_\alpha(\zeta_\alpha) &=& \lim_{p\to 0}
 \left[\zeta_\alpha + D_\alpha |p|^{2-z_\alpha}
 + Q_{\alpha\alpha}
  \zeta_\alpha^{2-z_\alpha -\frac{1}{z_\alpha}} |p|^{3-2 z_\alpha}\right.\nonumber \\
 &+& \left. \sum_{\beta\neq \alpha} Q_{\alpha\beta} \left(-i v_p^{\alpha\beta}
    \right)^{\frac{1}{z_\beta}-1}|p|^{1+\frac{1}{z_\beta}-z_\alpha}
\right]^{-1}.
\eea

with $v_p^{\alpha\beta} :=  |v_\alpha-v_\beta| \sign[p(v_\alpha-v_\beta)]$ and
\bel{Qab}
Q_{\alpha\beta} = 2 (G^{\alpha}_{\beta\beta})^2
\Gamma\left(1-\frac{1}{z_\beta}\right)\Omega[\hat{S}_\beta] \geq 0.
\ee
where
\bel{intf}
\Omega[\hat{f}] = \int_{-\infty}^\infty \rmd p\, \hat{f}(p)\hat{f}(-p).
\ee
With $\sigma_p^{\alpha\beta} = \sign[p(v_\alpha-v_\beta)]$ one has
\bea
& &\left(-i v_p^{\alpha\beta}\right)^{\frac{1}{z_\beta}-1}
 = \sin{ \left(\frac{\pi}{2z_\beta}\right)  }
|v_\alpha-v_\beta|^{\frac{1}{z_\beta}-1} \times
\nonumber \\
& &\left[1 - i \sigma_p^{\alpha\beta} \tan{\left( \left(1+\frac{1}{z_\beta}
 \right) \frac{\pi }{2}\right)}\right].
\label{ivab3}
\eea

Now we are in a position to analyze the small-$p$ behaviour.  One has
to search for dynamical exponents for which the limit $p\to 0$ is
non-trivial, which is determined by the smallest power of $p$ in
\eref{memoryLFT5}.  This has to be done self-consistently for all
modes. We find that different conditions arise depending on which
diagonal elements of the mode-coupling matrices vanish.  In order to
characterize the possible scenarios we define the set $\I_\alpha :=
\{\beta: G^{\alpha}_{\beta\beta} \neq 0\}$ of non-zero diagonal mode
coupling coefficients. One obtains from \eref{memoryLFT5} through
power counting the system of equations
\be
\label{dynexpsumm2a}
z_\alpha = \left\{ \ba{lll}
2 & \mbox{ if } & \I_\alpha = \emptyset  \\
3/2 & \mbox{ if } & \alpha \in \I_\alpha\\
\min_{\beta \in \I_\alpha}\left[\left(1 + \frac{1}{z_\beta}\right)\right]
& \mbox{ else } &
\ea \right.
\ee
and the domain
\be
\label{dynexpsumm1a}
1 < z_\alpha \leq 2 \quad \forall \alpha.
\ee
for the possible dynamical exponents.

\end{appendix}


\begin{acknowledgments}
We thank Herbert Spohn for helpful comments on a preliminary version of
the manuscript.
This work was supported by Deutsche Forschungsgemeinschaft (DFG)
under grant SCHA 636/8-1.
\end{acknowledgments}





\begin{thebibliography}{99}

\bibitem {GoldenRatioLivio} Livio M (2003) The Golden Ratio:
The Story of PHI, the World's Most Astonishing Number,
Broadway Books (ISBN 978076790815).

\bibitem{Land14}
Landi GT, de Oliveira MJ (2014)
Fourier's law from a chain of coupled planar harmonic oscillators
under energy-conserving noise.
Phys. Rev. E 89(2): 022105.

\bibitem{Gend14}
Gendelman OV, Savin AV (2014)
Normal heat conductivity in chains capable of dissociation.
EPL 106: 34004.

\bibitem{Kard86}
Kardar M, Parisi G, Zhang Y-C (1986)
Dynamic scaling of growing interfaces.
Phys. Rev. Lett. 56: 889--892.

\bibitem{Spoh14}
Spohn H (2014) Nonlinear Fluctuating hydrodynamics for anharmonic chains.
J. Stat. Phys. 154: 1191--1227.

\bibitem{Spohn15}
Spohn H (2015)
Fluctuating hydrodynamics approach to equilibrium time correlations
for anharmonic chains.
arXiv:1505.05987.

\bibitem{Maun97}
Maunuksela J, Myllys M, K\"ahk\"onen O-P, Timonen J, Provatas N,
Alava MJ, Ala-Nissila T (1997)
Kinetic roughening in slow combustion of paper.
Phys. Rev. Lett. 79: 1515.

\bibitem{Miet05}
Miettinen L, Myllys M, Merikoski J, Timonen J (2005)
Experimental determination of KPZ height-fluctuation distributions.
Eur. Phys. J. B 46: 55--60.

\bibitem{Waki97}
Wakita J, Itoh H, Matsuyama T, Matsushita M (1997)
{\em Self-affinity for the growing interface of bacterial colonies},
J. Phys. Soc. Jpn. 66(1): 6772.

\bibitem{Yunk13}
Yunker PJ, Lohr MA, Still T, Borodin A, Durian DJ, Yodh AG (2013)
Effects of particle shape on growth dynamics at edges of evaporating
drops of colloidal suspensions.
Phys. Rev. Lett. 110: 035501.


\bibitem{KPZniceIntroduction}
For a nice introduction into the
  KPZ class and its relevance we refer to \cite{Buchanan14}. Recent
  reviews \cite{HalpinHealy,QuastelSpohn} provide a more detailed
  account of theoretical and experimental work on the KPZ class.

\bibitem{Buchanan14}
Buchanan M (2014)
Equivalence principle.
Nature Physics 10(8): 543.

\bibitem{HalpinHealy}
Halpin-Healy T, Takeuchi K.A. (2015)
A KPZ cocktail - shaken, not stirred: toasting 30 years of kinetically
roughened surfaces.
J. Stat. Phys. 160: 794.

\bibitem{QuastelSpohn}
Quastel J, Spohn H (2015)
The one-dimensional KPZ equation and its universality class.
J. Stat. Phys. 160: 965.


\bibitem{Prae02}
Pr\"ahofer M, Spohn H (2004)
Exact scaling functions for one-dimensional stationary KPZ growth.
J. Stat. Phys. 115: 255--279.

\bibitem {Prae04}
Pr\"ahofer M, Spohn H (2002)
Current fluctuations in the totally asymmetric simple exclusion process.
\textit{In and Out of Equilibrium}, Vol.~51 of Progress in Probability,
ed. Sidoravicius V (Birkhauser, Boston).

\bibitem{Take10}
Takeuchi KA, Sano M (2010)
Universal fluctuations of growing interfaces: evidence in turbulent
liquid crystals.
Phys. Rev. Lett. 104: 230601.

\bibitem{Take11}
Takeuchi KA, Sano M, Sasamoto T, Spohn H (2011)
Growing interfaces uncover universal fluctuations behind scale invariance.
Sci. Rep. 1: 34.

\bibitem{vanB12}
Van Beijeren H (2012)
Exact results for anomalous transport in one-dimensional Hamiltonian systems.
Phys. Rev. Lett. 108: 108601.

\bibitem{Mend13}
Mendl  CB, Spohn H (2013)
Dynamic correlators of FPU chains and nonlinear fluctuating hydrodynamics.
Phys. Rev. Lett. 111: 230601.

\bibitem{Spoh14b}
Spohn H, Stoltz G (2015)
Nonlinear fluctuating hydrodynamics in one dimension: The case of
two conserved fields.
J. Stat. Phys. 160: 861.

\bibitem{Popk14}
Popkov V, Schmidt J, Sch\"utz GM (2014)
Superdiffusive modes in two-species driven diffusive systems.
Phys. Rev. Lett. 112: 200602.

\bibitem{Popk14b}
Popkov V, Schmidt J, Sch\"utz GM (2015)
Universality classes in two-component driven diffusive systems.
J. Stat. Phys. 160: 835--860.

\bibitem{Gris11}
Grisi R, Sch\"utz GM (2011)
Current symmetries for particle systems with several conservation laws.
J. Stat. Phys. 145: 1499--1512.

\bibitem{Toth03}
T\'oth B, Valk\'o B (2003)
Onsager relations and Eulerian hydrodynamic limit for systems
with several conservation laws.
J. Stat. Phys. 112: 497--521.

\bibitem{Devi92}
Devillard P, Spohn H (1992)
Universality class of interface growth with reflection symmetry.
J. Stat. Phys. 66: 1089--1099.

\bibitem{Durrett}
Durrett R (2010) Probability: Theory and Examples (4th Edition),
Cambridge University Press, Cambridge, p. 141.

\bibitem{Lee10}
Lee CH, Choi W, Han J-H, Strano MS (2010)
Coherence resonance in a single-walled carbon nanotube ion channel.
Science 329: 1320--1324.

\bibitem{Popk04}
Popkov V, Salerno M (2004)
Hydrodynamic limit of multichain driven diffusive models.
Phys. Rev. E 69: 046103.

\bibitem{Card96}
Cardy J (1996) \textit{Scaling and Renormalization in Statistical Physics}
(Cambridge University Press, Cambridge)

\bibitem{Henk99}
Henkel M (1999) {\em Conformal Invariance and Critical Phenomena}
(Springer, Berlin).

\bibitem{Henk02}
Henkel M (2002)
Phenomenology of local scale invariance: From conformal invariance
to dynamical scaling.
Nucl. Phys. B 641: 405--486.

\bibitem{Parisi}
Pagnani A, Parisi G (2015)
Numerical estimate of the Kardar-Parisi-Zhang universality class in
(2+1) dimensions.
Phys. Rev. E 92, 010101

\bibitem{HalpinHealy2}
Halpin-Healy, T (2012)
(2+1)-dimensional directed Polymer in a random medium: scaling phenomena
and universal distributions.
Phys. Rev. Lett. 109, 170602

\bibitem{Scaling2dKPZ}
Kloss T, Canet L, Wschebor N (2012)
Nonperturbative renormalization group for the stationary
Kardar-Parisi-Zhang equation: Scaling functions and amplitude ratios
in 1+1, 2+1, and 3+1 dimensions.
Phys. Rev. E86, 051124


\end{thebibliography}
\end{document}